\documentclass[conference,compsoc]{IEEEtran}
\usepackage{amsmath}
\usepackage{amsfonts}
\usepackage{amssymb}
\usepackage{hyperref}
\usepackage{subfig}

\usepackage{mathtools}
\newcommand{\defeq}{\vcentcolon=}
\ifCLASSOPTIONcompsoc
  \usepackage[nocompress]{cite}
\else
  \usepackage{cite}
\fi
\ifCLASSINFOpdf
\else
\fi
\begin{document}
%
\title{Training Transformers for Information Security Tasks:\\
A Case Study on Malicious URL Prediction}

\author{\IEEEauthorblockN{Ethan M. Rudd}
\IEEEauthorblockA{FireEye Data Science\\
FireEye Inc.\\
Reston, VA\\
Email: ethan.rudd@fireeye.com}
\and
\IEEEauthorblockN{Ahmed Abdallah}
\IEEEauthorblockA{FireEye Data Science\\
FireEye Inc.\\
Reston, VA\\
Email: ahmed.abdallah@fireeye.com}}


%


\maketitle

\begin{abstract}

Machine Learning (ML) for information security (InfoSec) utilizes distinct data types and formats which require different treatments during optimization/training on raw data. In this paper, we implement a 
malicious/benign URL predictor based on a transformer architecture that is trained from scratch. We show that in contrast to conventional natural language processing (NLP) transformers, this model requires a different training approach 
to work well. Specifically, we show that 1) pre-training on a massive corpus of unlabeled URL data for an auto-regressive task does not readily transfer to malicious/benign prediction but 2) that using an auxiliary auto-regressive loss improves performance when training from scratch. We introduce a method for mixed objective optimization, which dynamically balances contributions from both loss terms so that neither one of them dominates. We show that this method yields performance comparable to that of several top-performing benchmark classifiers.


\end{abstract}


%
\IEEEpeerreviewmaketitle

\section{Introduction}

The abundance of labeled data 
sources has been a major driver in the success of ML in applications across the board, starting with computer vision, and spreading to many other domains, including InfoSec. However, applications of ML to InfoSec often deal with unique data formats which must be approached differently. This is one of the reasons why, unlike computer vision, NLP, and audio applications to name a few -- which have also seen similar performance improvements and larger labeled data sets -- InfoSec applications of ML still rely heavily on relatively simple models fit on hand-tuned features.

This may be in part because properly learning features based on raw data is less trivial for some of the data formats inherent to security problems, for example, in \cite{anderson2018ember}, MalConv – a convolutional neural network which operates on raw PE files and is closely analogous to vision or NLP convnets \cite{raff2018malware} -- underperforms a lightGBM model trained with default parameters on hand-engineered features. 

Intuition suggests that hand-engineered features potentially leave substantial performance on the table. Given a suitable optimization technique a model capable of operating on raw data can derive its own features that are well-optimized for the task at hand. Moreover, employing such a model allows for a simpler train/deploy pipeline across different formats. Thus there is an impetus for research into more effective ML models for InfoSec tasks that operate on raw data.

With respect to other domains where ML models operating on raw data have advanced the state of the art, natural language processing (NLP) is arguably most analogous to InfoSec tasks. Some InfoSec tasks can even be framed as NLP problems, including email spam detection \cite{dada2019machine}, website content categorization \cite{rao2019detection}, social media disinformation triage \cite{seymour2016weaponizing}, and certain areas of data leak prevention (DLP) \cite{alzhrani2016automated,alzhrani2016automated2,alzhrani2017automated}. Other InfoSec tasks, which cannot directly be framed as NLP problems still face similar challenges, for example, these tasks typically contain long-ranging sequential dependencies, where, at inference, one or more tokens that are far-removed from a given token in the sequence may strongly influence the probability of that token. Given the close analogy to NLP and because transformer architectures have revolutionized performance in the NLP domain, it seems fruitful to explore the feasibility of applying transformers to InfoSec tasks, especially since they somewhat address the long range dependency issue. 

However, InfoSec is a broad field, and sequential dependency ranges vary dramatically based on the data type.  In NLP literature, “long-range” dependencies are typically 
considered to span tens to tens of thousands of tokens – often backwards -- within a sequence. While similar ranges may apply for certain types of InfoSec tasks – e.g., infosec tasks that are built on natural language – for other tasks -- e.g., binary classification where function calls and declarations can be almost arbitrarily separated -- relevant dependencies can span forward or backward millions or even billions of bytes.

While some research has been conducted on scaling transformer context window sizes and additional research is required to develop transformer architectures for other InfoSec tasks, in this paper, we perform an initial feasibility study in which we train transformers to detect malicious URLs. Malicious URL detection has the advantage that it is an InfoSec task which cannot be framed directly as an NLP problem, yet the relatively short sequence lengths fit into transformers' context windows without introducing additional modifications for longer sequences (e.g., hidden state caching, sparse attention patterns, locality sensitive hashing, etc.). 

Malicious/benign URL data/classification also has certain characteristics which, to a point, resemble those of other InfoSec datasets/problems: Contrary to NLP tasks, where large datasets are available for pre-training, but relatively small datasets are available for fine tuning, many InfoSec classifiers are fit on large quantities of weakly labeled data, labeled, e.g., by aggregating over threat feeds or externally derived vendor scores. Thus, this feasibility study provides some preliminary insight on how best to train transformer architectures in this problem setting. Contemporaneous work on scaling transformer context windows for different InfoSec data formats, we defer to another paper. The contributions of this paper are as follows:


\begin{itemize}
    
    \item A performance comparison between our novel transformer approaches and other more conventional approaches to URL classification. Using a transformer model, we are able to achieve performance on par with our top benchmark models.
    
    \item A performance comparison of multiple URL transformer training regimes, including with and without auto-regressive pre-training. We demonstrate that in contrast to NLP applications which rely heavily on unsupervised pre-training, this strategy does not readily improve performance on the URL classification task.
    
    \item A novel balanced mixed objective transformer loss function which balances a classification objective with an auxiliary auto-regressive objective during training.
    
 \end{itemize}

\section{Background}

\begin{figure*}[ht]
\centering
\subfloat[Next Character Prediction]{\includegraphics[width=0.3\textwidth]{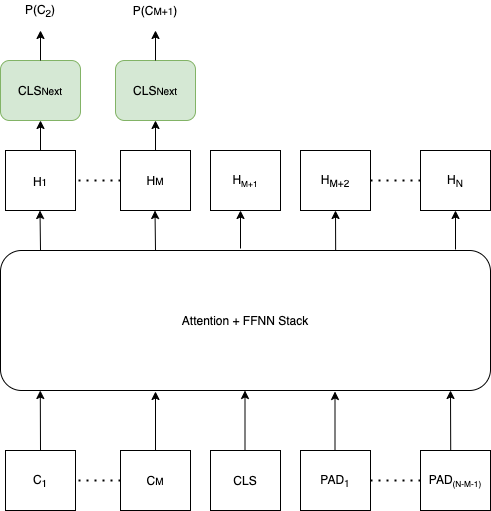}\label{fig:next}}\hfill
\subfloat[Classification Only]{\includegraphics[width=0.3\textwidth]{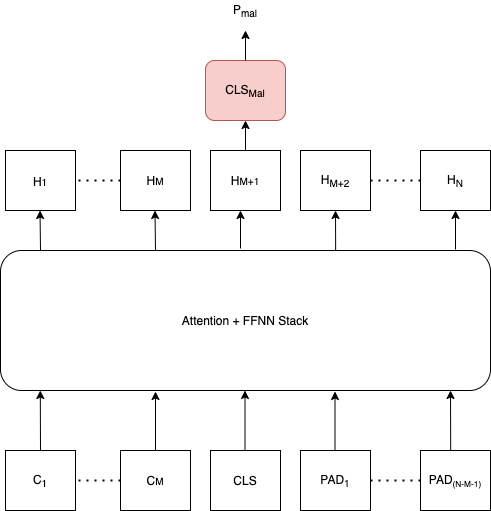}\label{fig:cls}}\hfill
\subfloat[Classification with Auxiliary Loss]{\includegraphics[width=0.3\textwidth]{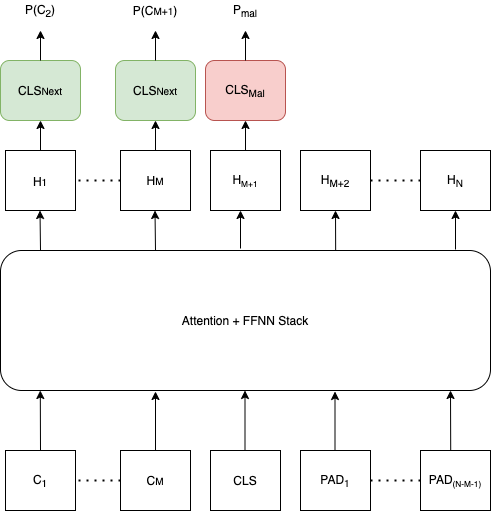}\label{fig:mixed}}
\caption{For a given context window size, transformers utilize padding/masking operations to perform inference over different sequence lengths. Let $M$ be the length of the sequence and $N$ be the size of the context window, where $M \leq N-1$. Let $C_1$ through $C_M$ be the embeddings corresponding to the sequence elements, in this case, characters within a URL. Let CLS be the embedding corresponding to the CLS token (position $M+1$), and let $PAD_1$ through $PAD_{N-M-1}$ be embeddings of padding tokens. Let $H_1$ through $H_N$ be the final hidden states derived from the transformer. A standard \textit{next character prediction} task is shown in  \protect\subref{fig:next}, where the hidden states are each passed through a classifier $CLS_{Next}$, and the softmax outputs correspond to probabilities of the subsequent tokens. \protect\subref{fig:cls} We can also use the output of $H_{M+1}$, passed through a binary classifier to derive a probability that the input sequence is malicious or benign. \protect\subref{fig:mixed} During training, we can use all token outputs to fit a malicious/benign predictor and a next-character prediction loss simultaneously. Note that hidden states corresponding to padding are ignored. For our implementation, L-R masking operations prevent prior hidden states from attending to any padding-related sequence information.}
\label{fig:transfomer}
\end{figure*}

ML for InfoSec has been researched for decades \cite{rudd2016survey}, but widespread industry adoption has occurred only over the past several years. Of the adopted models, the majority of them utilize some form of hand-crafted features (e.g., \cite{rudd2018meade,rudd2019aloha,anderson2018ember,raff2019kilograms,kyadige2019learning,ducau2019smart}), with only a few operating on raw data. Sequence models, e.g., recurrent neural networks (RNNs) have been applied with some success \cite{pascanu2015malware}, but only to niche problems. Training RNNs for most problems is fundamentally not scalable to long sequences, both in terms of the incremental training times and in terms of lack of the general loss of information from long term dependencies.

Transformers, introduced by Vaswani et al. in \cite{vaswani2017attention} have revolutionized natural language processsing (NLP) and a variety of other discrete sequence modeling domains. This is partly due to their ability to directly incorporate long-term dependencies in a sequence and partly because they are easily trained in parallel. Transformers were originally formulated for sequence transduction tasks, e.g., neural machine translation (NMT), wherein a source sequence is encoded via an encoder stack of attention and feed-forward layers, then decoded via a similar decoder stack, using the input sequence as a source context for decoding to the target sequence. E.g., in an English to French translation model, the encoding of the English sentence would serve as context for the French translated sentence; during translation, the context encoding along with hidden states from the translated portion of the sentence are used to arrive at the next token value. Later models, which constitute massively pre-trained representations, meant for fine-tuning for non-transductive tasks abandon the encoder/decoder setup entirely, opting in favor of either an encoder or a decoder. 

Bidirectional models, e.g., BERT\cite{devlin2018bert}, leverage masked language modeling pre-training to arrive at an encoder whose hidden states can be used for a variety of NLP fine tuning tasks. OpenAI's GPT \cite{radford2018improving} series of models apply a similar approach, but leverage left to right (L-R) masked modeling, in a next-token prediction training regime, wherein a decoder is used to predict subsequent elements of the sequence from previously decoded tokens. A wide variety of modifications to transformers have been made as well, including modeling over multiple modalities and memory/processing optimization tricks to extend context windows to longer lengths, e.g., \cite{lu2019vilbert,sukhbaatar2019adaptive,beltagy2020longformer,kitaev2020reformer,dai2019transformer}. The latter approaches could be interesting when applying transformers to other InfoSec problems/data formats where longer sequences are typical, e.g., classifying malware/benignware from raw bytes of a PE file. In this work, however, we work with relatively short sequences and examine fundamental transformer modeling approaches for a typical InfoSec task. 

In contrast to NLP problems where massive corpora are available for pre-training a generic representation, but potentially few labeled examples are available for fine-tuning to a task of interest, many InfoSec applications where ML works well have millions to hundreds of millions of labeled samples. These labels are typically derived from an aggregation of multiple weak labeling sources \cite{ratner2017snorkel,rudd2018meade,fu2020fast}. 

We are not the first to apply transformers to InfoSec tasks. Li et al. applied transformers to malware detection in \cite{li2019mad}. However, contrary to their approach which is a hierarchical system with multiple transformers trained with multiple different training/fine-tuning regimes, our approach is built on a single end-to-end transformer. While their work is interesting, their evaluation is severely limited in scale ($\sim$ 10k samples) and further evaluation is needed to assess its efficacy in realistic scenarios.

Intriguingly, transformers are amenable to multi-target training, and those that perform next character prediction do so implicitly. Our balanced training methodology, however, is a \textit{multi-task} approach \cite{rudd2016moon}. Several works \cite{rudd2019aloha,ducau2019automatic,huang2016mtnet} have shown that similar multi-task optimization improves the performance of ML on InfoSec tasks. In contrast to these previous approaches, however, it uses a novel technique for re-weighting the per-task loss, such that no auxiliary loss term ever dominates regardless of the loss magnitude.

Our work is not the first to address malicious URL detection -- there are several approaches in the literature that use convolutional architectures to classify malicious/benign URLs and DGA-generated domains including \cite{sahoo2017malicious,saxe2017expose,le2018urlnet,yu2018character}. While these works are interesting, note that they use different benchmark data sets, and are thus not directly comparable to our work. We benchmark against similar convolutional and non-convolutional architectures in this paper, which we have developed internally. Note that the primary focus of this work is not creating the optimal URL classifier, but exploring how to train transformers for a typical non-NLP InfoSec machine learning task and contrasting with transformer training regimes for NLP problems.

While transformers can be implemented in a variety of ways, a typical implementation of a transformer attention stack consists of the following components: 

\begin{enumerate}
    \item Embedding: The input, in our case a URL, is projected into an embedding space. 
    \item Positional Encoding: This is a signal added to each of the embedding vectors to imbue each embedding with a positional order in the sequence (which would otherwise not be tracked by the attention mechanism). This can optionally be added to hidden states within the transformer, but we do not do this.
    \item Multi-Headed Attention Layers: These apply the attention mechanism in parallel in a redundant fashion across $h$ heads. The results are concatenated and fed to a feed-forward layer.
    \item Feed-Forward Layers: These act on each of the hidden states produced by multi-headed attention.
    \item Residual Connections: These add and norm operation between inputs to each stack of attention/FFNN layers. This allows information to percolate up the network and bypass the attention/FFNN blocks when appropriate.
    \item Masks: Because transformers operate in parallel, padding is typically added to inputs shorter than the context window of the transformer. In order to avoid ``attending" to the padding, masking is used. Masking is also used to enforce sequential dependencies, e.g.,  for our implementation, we enforce an L-R sequential dependence.
\end{enumerate}

We will elaborate on the architecture of our specific implementation later on in this paper. However, we present a high-level schematic of how multiple prediction tasks can be performed using a transformer in Fig. \ref{fig:transfomer}. Fig. \ref{fig:next} depicts a standard next-character prediction task from the literature. In this case, the outputs are softmax probabilities over the vocabulary. Fig. \ref{fig:cls} depicts using the transformer as a binary malicious/benign classifier -- the baseline ``decode-to-label" approach presented in Sec. \ref{sec:baseline}. Fig. \ref{fig:mixed} depicts the mixed objective approach presented in Sec. \ref{sec:mixed}, where a loss over the next character prediction output is used as an auxiliary loss in conjunction with the main classification task loss during training.

\section{Approach}

Let $X$ be a dataset of URLs with binary labels $Y$. Let $x$ be a generic URL from $X$ with label $y \in \{0,1\}$, 0 corresponding to benign and 1 corresponding to malicious. 

\subsection{Baseline: Decode-to-Label}
\label{sec:baseline}

Our ``decode-to-label" approach uses a left to right (L-R) decoder, similar to OpenAI's GPT \cite{radford2018improving} approaches, with a $\langle$CLS$\rangle$ token placed at the end of the sequence. Sequence information is propagated through the attention layer states in a L-R manner, meaning that hidden state $k$ in the $l$th attention layer is fed information corresponding to hidden states $1,2,\hdots,k-1,k$ in the $(l-1)$th layer. The final classification is made via a feed-forward neural network (FFNN), which is fed the top layer's hidden state corresponding to the $\langle$CLS$\rangle$ token. The final dense layer of the FFNN projects the output to a 1D value. This value is then passed through a Sigmoid to assume a final prediction  $h(x) \in [0,1]$. Binary cross entropy between the prediction $h(x)$ and label $y$ is evaluated during training and the associated gradients are backpropagated. The associated \textit{Classification Loss} is then:

\begin{equation}
    L_{CLS}(x,y) = -y log(h(x)) + (1-y)log(1-h(x)).
\end{equation}

In contradistinction to most transformer literature, this approach does not utilize any explicit loss over hidden states corresponding to tokens within the sequence. Our rationale for applying the ``decode-to-label" approach is to provide a benchmark against which to assess any gains/losses introduced explicitly optimizing sequential information into the transformer. 

\subsection{Next Character Prediction Pre-Training and Fine-Tuning}
\label{sec:next}

Next character prediction tasks have been applied throughout the transformer literature. For this regime, no labels are used during pre-training. Instead, we have a pre-training set $\hat{X}$ which may or may not have associated malicious/benign labels. Each subsequent character in the URL serves to label each previous character. For example, given input URL

\begin{center}
$\hat{x}=$\url{https://www.fireeye.com/},
\end{center}

\noindent
 the ``label" sequence would be

\begin{center} 
\url{ttps://www.fireeye.com/}$\langle$CLS$\rangle$. 
\end{center}
 
Note that the $\langle$CLS$\rangle$ token is omitted from the input sequence. 

We encode each character with its respective ASCII byte value ranging from 0 to 255. Note that in practice only a subset of these byte values are manifest in our data. We use the value 256 to represent our $\langle$CLS$\rangle$ token, yielding 257 distinct input embeddings.

Next character prediction is performed over each hidden state of the transformer output, corresponding to the embedded vector of the sequence up until that point. A feed-forward neural network (FFNN), which takes the corresponding transformer hidden state is used as a predictor of the next character. The output of the FFNN is a 257-element softmax, with the first 256 output probabilities corresponding to the probabilities of specific byte values as the next character and the last output probability corresponding to the probability of the $\langle$CLS$\rangle$ token (i.e., the end of the sequence). Note that for the next character prediction task, the input sequence $\hat{x}$ ignores the $\langle$CLS$\rangle$ token at input; this token is only used as a ``label" for the last character of the URL.

Loss is evaluated as the categorical cross entropy over the entire sequence, normalized by the sequence length $M$. Let $I(\cdot)$ be an indicator function which evaluates to 1 if the argument is true and 0 otherwise. The next character loss function over the URL $\hat{x}$ becomes:

\begin{equation}
L_{NEXT}(\hat{x}) = -\frac{1}{M}\sum_{i=1}^{M}\sum_{j=0}^{256} I(\hat{x}_{i+1} = j) log (h(\hat{x}_i)_j).
\end{equation}

Backpropagation of $L_{NEXT}$ is used to train the underlying transformer representation. During fine-tuning, the pre-trained representation is loaded, potentially with lower attention layers frozen. With a decreased learning rate, $X$ and $Y$ are fed to the transformer with $L_{CLS}$ used to train a malicious/benign predictor and update the unfrozen parameters of the transformer.  

\subsection{Balanced Mixed Objective Training}
\label{sec:mixed}

This approach aims to jointly optimize for both next character prediction and malicious/benign classification across dataset $X$ with labels $Y$. Note that this does not preclude next character prediction pre-training over another dataset. The rationale behind this approach is built on prior research which suggests that optimizing over multiple (correlated) tasks simultaneously leads to a better performing classifier with more stable convergence characteristics.

Following this rationale, we apply a mixed objective optimization approach, which balances main malicious/benign determination task with an auxiliary next character prediction loss. Contrary to previous research, which uses ad-hoc fixed weights on main and auxiliary task losses, our novel approach employs an adaptive balancing scheme, which ensures that no single loss term dominates, regardless of the loss value. Our loss-weighting strategy is as follows: 
 
\begin{align}
\begin{split}
    L_{MIXED_I} = &\alpha_I L_{CLS_I}+ \beta_I L_{NEXT_I}.
\end{split}
\end{align}
\noindent

At iteration $I$ of training, corresponding to one mini-batch, values $\alpha_I$ and $\beta_I$ are balancing multipliers computed for each mini-batch, and are assumed constant when computing the gradient of the loss function. They are used to weight the classification and next character prediction loss components.
Respectively, $\alpha_I$ ensures that $L_{next_I}$ accounts for $\frac{a}{a+b}$ of $L_{mixed_I}$ and $\beta_I$ ensures that $L_{cls_I}$ accounts for $\frac{b}{a+b}$ of $L_{mixed_I}$. Values $a$ and $b$ are hyperparameters which we fix during training.
Note that for simplification we can normalize such that $a+b:=1$ and say that $a$ and $b$ are the respective loss fractions themselves.  For our experiments, we set $a:=b:=0.5$ unless specified otherwise. Values $\alpha_I$ and $\beta_I$ are computed for each minibatch, according to:


\begin{align}
    \alpha_I &= \frac{a(L_{CLS_I}+L_{NEXT_I})}{L_{CLS_I}},\\
    \beta_I &= \frac{b(L_{CLS_I}+L_{NEXT_I})}{L_{NEXT_I}}.
\end{align}

Note that in this work, we apply only two loss types. However, we could trivially extend our approach to $K$ different loss types as follows. Given generic multiplier $\gamma_{iI}$ for the $i$th loss term at iteration $I$, $L_{iI}$, and desired loss contribution fraction $c_i/\sum_{j=1}^K c_j$, we compute $\gamma_{iI}$ as follows: 

\begin{equation}
\gamma_{iI} = \frac{c_i\sum_{j=1}^K L_{jI}}{L_{iI}}.
\end{equation}

\section{Experiments}

We collected a dataset of URLs, down-sampled over 2-3 months from a threat intel feed in late 2019. Each URL had multiple weak malicious/benign labels. We derived a single malicious/benign label for each URL using Snorkel \cite{ratner2017snorkel}. The training set consisted of 1,007,451 labeled URLs with 180,052 malicious and 826,333 benign. The validation set consisted of 111,930 URLs with 20,013 malicious and 91,813 benign. The test set consisted of 279,874 URLs with 50,029 malicious and 229,604 benign respectively. We additionally collected a dataset of 20 million unlabeled URLs for pre-training experiments, disjoint from train, test, and validation sets. Note that we did not perform any cleanup of these URLs, meaning that our dataset includes lengthy/obscure URLs as well as raw IPs; thus the reported performance numbers are not comparable with those from other authors with proprietary datasets (e.g., \cite{saxe2017expose}).

\subsection{Transformer Base Topology}

We implemented a transformer with 20 layers of attention, a context window size of 256, a model hidden state size of 64, a feed-forward dimension of 128, and 4 attention heads per-layer. We also employed dropout, with a dropout ratio of 0.1. For classification, we apply a feed forward neural network (FFNN) atop the output corresponding to the $\langle$CLS$\rangle$ token. This FFNN consists of a 64-dimensional input followed by Layer Normalization, a hidden layer of 32-dimensions, Exponential Linear Unit activation, another hidden layer, reducing dimensionality from 32 to 1, and a sigmoid output. For next character prediction, we apply a similar architecture, but with a softmax output of 257 dimensions, predicting bytes 0 through 255 and the $\langle$CLS$\rangle$ token (which we ascribe label 256).

\subsection{Comparison of Different Training Regimes}

We performed comparisons of our four training techniques, ensuring that performance converged on the validation set. 

{\bf Baseline (DecodeToLabel):} As a baseline, we compute loss only over the binary malicious/benign prediction, performing neither pre-training nor next character prediction. We trained the model for 15 epochs with minibatch size 512.  For this training regime we used PyTorch's default Adam optimizer.

{\bf Auto-regressive Pre-training and Fine Tuning -- Training Set (FineTune)} For this experiment, we first performed pre-training over the entire training set for the next-character prediction task for 15 epochs. We then performed fine-tuning, freezing the first 16 attention layers and trained the decode-to-label task for 15 epochs at a reduced learning rate. For Pre-Training, we used PyTorch's default Adam Optimizer. For fine-tuning, we used SGD with a learning rate of 1e-4.

{\bf Auto-regressive Pre-training and Fine Tuning -- Pre-Training Set (FineTune 20M)} For this experiment, we first performed pre-training over the 20 million URL pre-training set for 2 epochs, using PyTorch's default Adam optimizer. We then performed fine-tuning, freezing the first 16 attention layers and trained the decode-to-label task for 15 epochs at a reduced learning rate. For fine-tuning, we used SGD with a learning rate of 1e-4.

{\bf Balanced Mixed Objective Training (MixedObjective)} We performed 15 epochs of balanced  mixed objective training over the training set using PyTorch's default Adam optimizer.

We report results in terms of Receiver Operating Characteristics (ROC) curves and area under the ROC curve (AUC). 

\begin{figure}[!h]
  \centering
      \includegraphics[width=\linewidth]{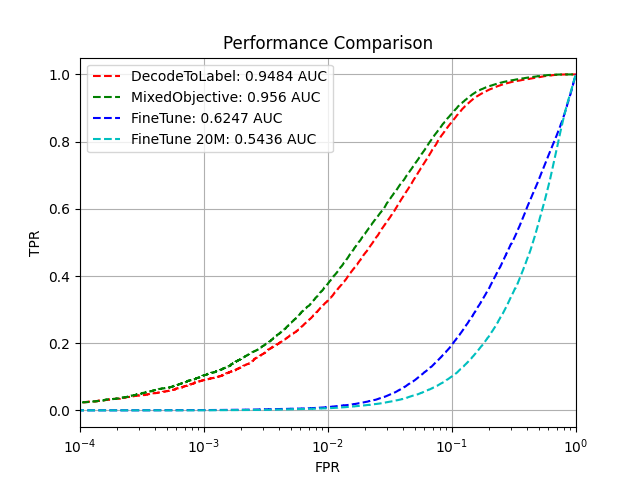}
  \caption{ROC and ROC-AUC results from different URL transformer training approaches. The Mixed Objective approach, shown in green, outperforms a generic decode-to-label approach. Neither fine-tuning approaches yielded impressive performance.\label{fig:PerformanceComparison}}
\end{figure}

Surprisingly, fine-tuning from a pre-trained initialization resulted in poor performance compared to training from scratch from a Xavier initialization (see Fig. \ref{fig:PerformanceComparison}). We additionally attempted fine-tuning, freezing all but the last 4 layers of the transformer, but witnessed similar performance to our original fine-tuning regime.  This is despite a consistent decrease and general convergence in loss for both of the pre-trained representations. This suggests that there is less immediate task transfer between next-character prediction for URLs than there is for next-character prediction in NLP contexts. 

Despite the failure of pre-training with an auto-regressive loss to deliver performance gains, we did find that we were able to achieve marginal performance improvements by introducing an auxiliary next-character prediction loss. This is consistent with the findings of Rudd et al. \cite{rudd2019aloha}.

\subsection{Comparison Models}

As a viability comparison, we benchmarked our transformer against several other models which we have developed for malicious/benign URL detection. Substantial development and testing effort went into these models. We summarize them in this section.

{\bf Random Forest on SME-Derived Features} For this model, proprietary features were derived with the help of subject matter experts (SMEs). These feature vectors consist of binary values/counts derived from parsing the URL and checking if specific parsed values from the URL string reside in various lists corresponding to likely indicators of malicious or benign content. The derived feature vectors therefore characterize the content of the URL. 

We then fit a random forest on these extracted feature vectors. The random forest classifier consisted of a 30 trees, each with a maximum depth of 20. Nodes were split based on information gain. For all other parameters, we used Scikit-learn's \cite{pedregosa2011scikit} defaults.

{\bf LSTM on Raw URLs} For this model, we fit a long short-term memory neural network (LSTM) over the URLs, using embeddings of size 50 and an LSTM hidden size of 100. We performed optimization for a max of 100 epochs with early stopping based on validation performance. During optimization, we used an Adam optimizer with Keras's default parameters and a minibatch size of 128.

{\bf 1D CNN on Raw URLs} This model first embeds each input token into a 40-dimensional vector, then follows up with a dropout layer (p=0.2) and a stack of two 1D convolutions with and ReLU activations. The convolutional layers have 256 and 100 filters respectively with kernel sizes 5 and 3. The output of the convolutional layer stack is then globally max-pooled and passed through a hidden size layer of size 256, a dropout layer (p=0.2), and a ReLU activation. This is then transformed to 1D via a final fully-connected layer and passed through a Sigmoid output.


\begin{figure}[!t]
  \centering
      \includegraphics[width=\linewidth]{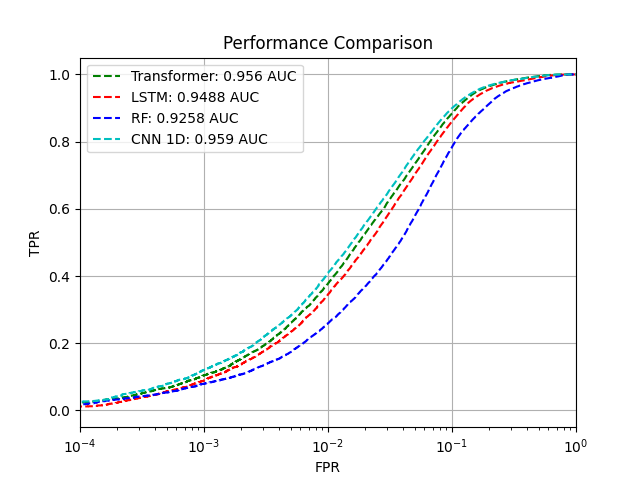}
  \caption{ROC and ROC-AUC results for comparative baselines along with our top performing transformer model. Note that the Mixed Objective transformer approach performs on-par with the highest performing baseline (CNN 1D), outperforming all other approaches.\label{fig:URLBaselines}}
\end{figure}

Our mixed objective transformer model (MixedObjective in Fig. \ref{fig:URLBaselines}) outperforms all but one of the comparison models, substantially outperforming the random forest feature-based model (RF) and the LSTM and performing on-par with the CNN model. 


\section{Discussion}

We have demonstrated that transformers can achieve performance comparable to or better than that of other top-performing models for URL classification. We have also found that, contrary to NLP domains, wherein auto-regressive pre-training substantially enhances performance in a fine-tuned regime, for our URL data, auto-regressive pre-training on a large corpus yields no apparent gains for the classification task and makes it substantially more difficult to fit a performant model. This suggests that the next character prediction task has too little apparent correlation with the task of malicious/benign prediction for effective/stable transfer. Interestingly, utilizing next character prediction as an auxiliary loss function \cite{rudd2019aloha} yields improvements over training solely to predict the label. Note that this occurs even with a relatively large portion of the overall loss term (50 \%) devoted to the auxiliary loss. This suggests that mixed objective optimization is more effective at correlating across heterogeneous loss terms than fine-tuning.

Note that in contrast to some of the comparison models we did not perform a rigorous hyperaparameter search for our transformer, since this research was primarily concerned with loss functions and training regimes; not an ``optimal" model topology. We leave this to future work, but the relatively impressive performance obtained herein suggests that  transformers could likely achieve state-of-the-art on the URL prediction.

\subsection{Model Interpretability Analysis}

We have shown that our URL transformer is capable of yielding competitive performance, but on what basis does it make its decisions? Ideally, we would like it to be detecting malicious URLs based on on potentially malicious content as opposed to spurious correlations. To get a qualitative idea of what the model is learning, we applied the method of integrated gradients\cite{sundararajan2017axiomatic} to our top performing transformer model for a sub-selection of malicious and benign URLs. The method of integrated gradients approximates the path integral -- along a linear path between a ``baseline" input and a sample -- of gradients of the score with respect to the input. This yields an element-wise view of the contributions of each element in the input to the overall score, the sum of which converges to the difference in scores between the sample and the baseline. In mathematical form,

\begin{equation}
    \text{IntegratedGrad}_i(x) \defeq (x_i - x_i^{'}) \int_{\alpha=0}^{1} \frac{\partial F{x_i^{'}} + \alpha(x - x^{'})}{\partial{x_i}} \partial \alpha
\end{equation}

\noindent
is the contribution of the $i$th character in the URL. For the baseline, we use a non-existent URL representation, corresponding to a zero vector in the embedded space feeding into the transformer.

Empirically, we observed the following:

\begin{itemize}
    \item Malicious classifications tend to occur based on high-entropy character strings.
    
    \item Semantic aspects of the url/domain like ``https" and ``.com" have little contribution to either malicious or benign classifications.
    
    \item Natural language strings embedded in the URL typically have a benign contribution to the overall score. 
    
\end{itemize}

These findings suggest that our URL transformer has managed to at least learn high-level intuitive features corresponding to whether a URL is malicious or benign.

\section{Conclusion}

We have demonstrated first steps for training transformers from scratch and applying them to an InfoSec dataset. While our URL dataset is not representative of all data in the InfoSec space, it does reflect the common property of having a multitude of samples with labels derived from weak labeling sources, differing from common NLP classification tasks. We have shown that for our use case, unlike in NLP domains, there is little benefit to pre-training, but substantial benefit from mixed objective optimization. We have introduced a novel loss function which dynamically re-balances gradients of auxiliary losses with the main task loss at each training step, potentially improving training stability. We are currently working to extend our models to other data formats, including formats which contain long sequences of generic bytes, to the tune of 1MB. This involves using some of the latest research in extended context sequence length. We are also investigating the use of bidirectional models with different training regimes.



\ifCLASSOPTIONcompsoc
  \section*{Acknowledgments}
\else
  \section*{Acknowledgment}
\fi

This research was funded by FireEye Inc. FireEye is a world-renowned cybersecurity company that specializes in cyber threat intelligence, cybersecurity data science, and cyber analytics.



\bibliographystyle{IEEEtran}
\bibliography{transformers}
%



\end{document}